\documentclass[11pt]{article}
\usepackage{moriond,epsfig}

\def\be{\begin{equation}}
\def\ee{\end{equation}}
\def\bea{\begin{eqnarray}}
\def\eea{\end{eqnarray}}

\begin{document}
\vspace*{4cm}
\title{Charm physics at the Tevatron}

\author{Mario Campanelli\footnote{On behalf of the CDF collaboration}}

\address{DPNC, Universit\'e de Gen\`eve ,\\
24, Q E. Ansermet Gen\`eve, Switzerland}

\maketitle\abstracts{The first period of the Tevatron Run II has been dominated
by low-Pt physics. Both detectors, and their trigger/DAQ systems have been
largely modified to exploit as much as possible the large wealth of charm
and beauty hadrons produced by the machine, with improvements in the tracking
system, magnetic field, specific triggers based on on-line track reconstruction.
Thanks to these improvements, Tevatron experiments are competitive with
dedicated machines, not only for heavier beauty states but also for charm,
both in leptonic and hadronic decays.
}

\section*{Introduction: why charm physics at the Tevatron}
CDF and D0 are known to the physics community for having discovered the
top quark, and being two experiments running at the world's energy frontier.
It is therefore not obvious that it makes sense for Tevatron experiments to 
invest a large amount of people and resources in sectors like physics of b and
c quarks, where also dedicated experiments are present, like the b-factories,
FOCUS and CLEO III.\par
The main reason why low-Pt physics is attractive at the Tevatron is the very
high production cross-section, allowing the detectors to record huge 
(in many cases, the world's largest) samples of charm and beauty hadrons.
As an example, the $b\bar{b}$ production cross section is about 50 $\mu$b at
the Tevatron compared to about 1 nb at the $e^+e^-$ b-factories. 
In some cases, like the heavy mesons $B_s$ and $B_c$, or the $\Lambda_b$,
the Tevatron is the only running accelerator where these states are actually
produced at all. However, the luminosity is a factor 1000 smaller than
at the b-factories, the detectors are optimized for higher energies, so
calorimetry and particle identification are not as good for low-Pt, and even
if these events are much cleaner than high-Pt cases like top production,
they still have far larger combinatorics compared to $e^+e^-$ colliders.

\section*{Accelerator and detector issues}
Tevatron Run II started in spring 2001 after several years of shutdown.
With respect to the previous run, the increase in CoM energy is modest (from
1.8 to 1.96 TeV), but the number of bunches in the machine has increased by
almost a factor 10 (timing between bunches from 3500 to 396 ns), to reach
instantaneous luminosities of $2\times 10^{32} cm^2s^{-1}$, and integrated
luminosities now expected between 4 and 9 $fb^{-1}$ before the first physics
results from the LHC. At the moment of writing (May 2004) Tevatron has 
delivered about 500 $pb^{-1}$ of data; the results presented in the following
are based on about 200 $pb^{-1}$ of data on tape.\par
Both detectors underwent major changes to cope with the new running conditions.
With 10 times shorter interval between bunch crossing, the full trigger and
data acquisition system had to be redesigned. Also the tracking systems
have been upgraded: both detectors have a new silicon detector, CDF has
replaced its old CTC with the similar but faster COT, while D0 has a radically
new tracking system, with central solenoidal magnetic field, a scintillating
fiber tracker and a preshower. Both detectors have also installed scintillators
to be used as TOF systems for particle identification and cosmic rejection.\par
Trigger issues are of course essential in a hadron collider. At a luminosity
of $10^{32} cm^2 s^{-1}$, production of b with $P_t>6$ GeV in the central
region has a rate of 1 kHz, while QCD events are produced at a rate of 1.7 MHz.
Events can only be written on tape at rates of 50-70 Hz. Triggering on 
heavy flavor events can be either done in the ``traditional'' RunI way, using
dilepton sample mainly from $J/\psi$ production, or, so far only in CDF,
using the new on-line track reconstruction provided by the SVT\cite{svt} 
tracking
system. The level-1 eXtremely Fast trigger (XFT) measures the curvature for
tracks with $P_t> 1.5$ GeV with a precision of $\sigma(P_t) = 1.74 \% P_t$;
this information is directly used for the $J/\Psi$ mass reconstruction
in the dimuon trigger, then passed to the SVT, where it is combined with
silicon hits to provide a measurement of all track parameters, including the
impact parameter with a resolution of 55 $\mu$m (out of which 33 are due to the
beam spread). 

\section*{Production and branching ratio}
The information from the SVT allowed CDF to design a Two-Track Trigger (TTT)
to study fully hadronic decays of beauty and charm. Requiring two tracks with
$P_t>2$ GeV, $d_0>100\mu$m, $\Sigma P_t> 5.5$ GeV, a very large $D^0$ sample
can be recorded, that can be made very pure by requiring an additional
``slow'' pion from $D^{*+}$ decay, with small $M(D*)-M(D^0)$ difference.\par
\begin{figure}
\begin{center}
\psfig{figure=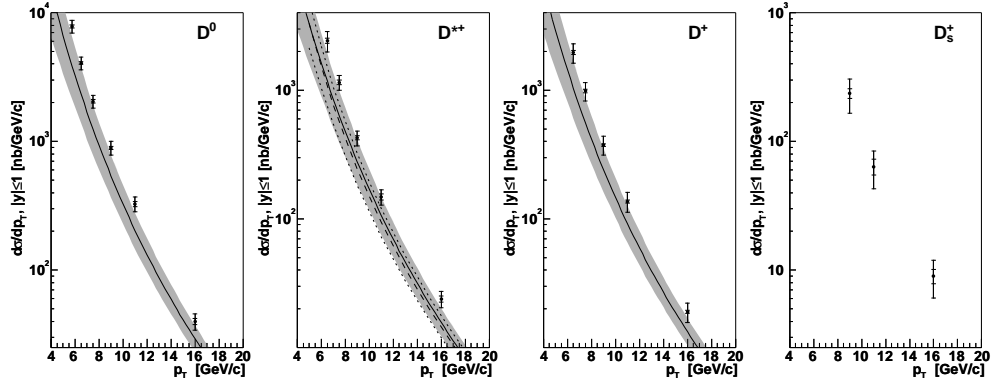,height=5cm}
\caption{Pt distribution of production cross section for of different charmed 
mesons
\label{fig:xsect}}
\end{center}
\end{figure}
CDF has published the cross section for $D^0$, $D^+$, $D^{*+}$ and $D_s$ 
production \cite{xsec},
shown in figure \ref{fig:xsect} to be in agreement with the upper side 
of the theory prediction in \cite{matteo}. From a fit to the 
distribution of the impact parameter 
of the D meson, the fraction of promptly produced D in a sample that also
contains events from b decay is $f_D = 86.6\pm0.4\pm3.5\%$.\par
The large $D^0$ sample is also used to measure relative branching ratios
and search for CP asymmetries. The relative branching fractions are found
to be $\Gamma(D^0\to K^+K^-)/\Gamma(D^0\to K\pi) = 9.96\pm 0.11\pm 0.12 \%$
$\Gamma(D^0\to \pi^+\pi^-)/\Gamma(D^0\to K\pi) = 3.608\pm 0.054\pm 0.040 \%$,
$\Gamma(D^0\to K^+K^-)/\Gamma(D^0\to \pi^+\pi^-) = 2.762\pm 0.040\pm 0.034 \%$
where in all of the above the first error is statistical and the second 
systematic. CP violation would show up in this sector as an symmetry between
decays of $D^0$ and $\bar{D^0}$ into $K^+ K^-$ and $K^- K^+$, respectively
(the same would be true for decays into pions). These measurements are better
than the previous best result from FOCUS \cite{focus}.\par
The CKM description of CP
violation leads to very small and practically immeasurable effects, in
agreement with the CDF measurements of $A(D^0\to K^+ K^-) = 2.0 \pm 1.2 \pm 
0.6$ and $A(D^0\to \pi^+ \pi^-) = 1.0 \pm 1.3 \pm 0.6$.

\section*{Spectroscopy}
The first paper published by CDF Run II \cite{dpmassdiff} was the mass 
difference between $D_s^+$ and $D^+$. Both states decay in $\Phi \pi$, with
subsequent $\Phi\to KK$ decay. What is measured is the difference between the
two peaks in the $\pi KK$ final state. Since this is a measurement of very
high accuracy, a very careful calibration of the tracking is needed, and the
bias induced by the fit has to be removed. After checking that some reference
resonances like $J/\Psi$ and $D^0$ had their mass at the PDG value and stable
as a function of $P_t$, the mass difference was measured to be
$M(D^+_s)-M(D^+) = 99.41 \pm 0.38 \pm 0,21$ MeV.\par

\begin{figure}
\begin{center}
\psfig{figure=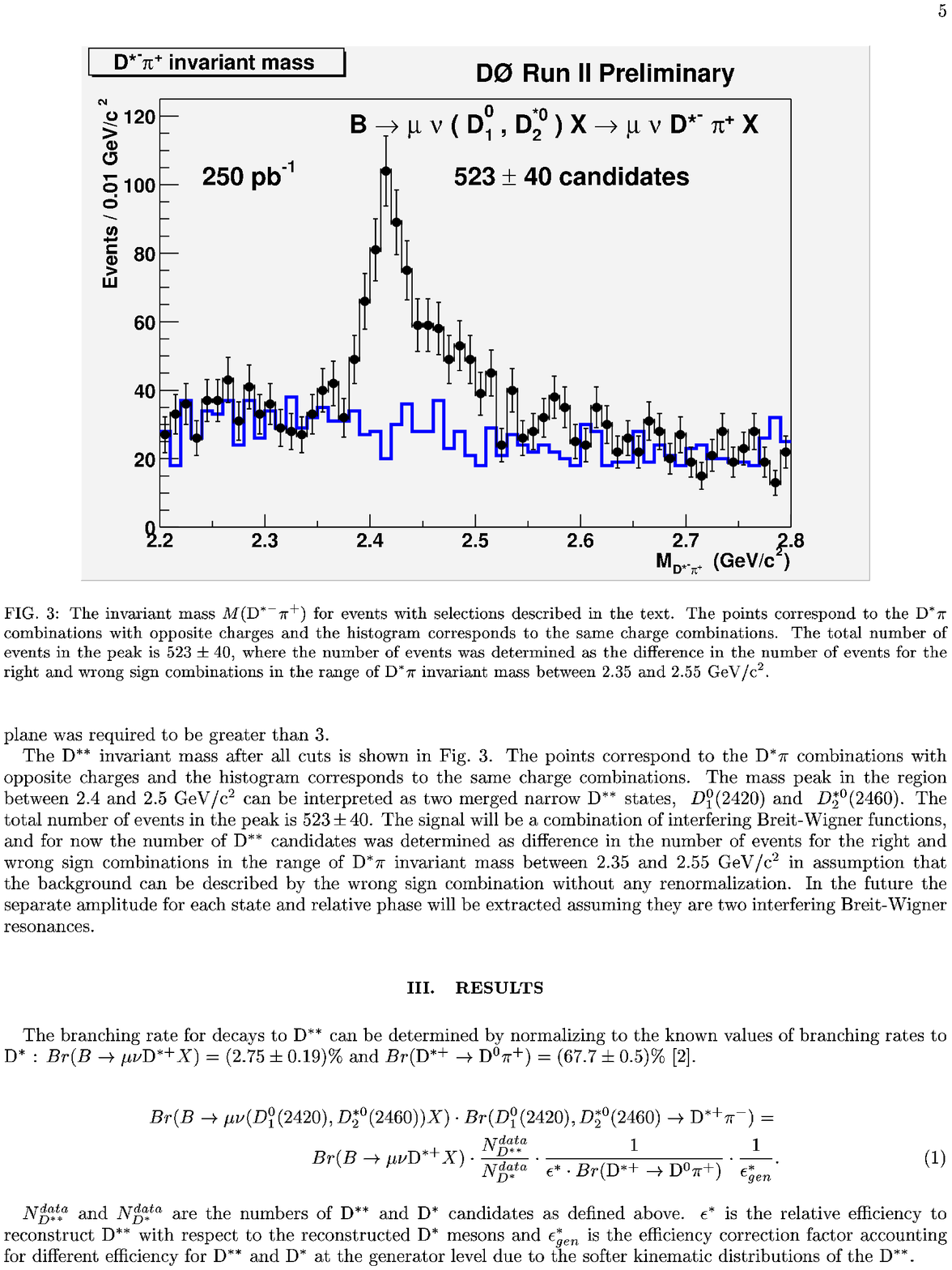,bb=95 414 520 700,height=5cm,clip}
\caption{Invariant mass of the two narrow $D^{**}$ mesons from semi leptonic
B decays.
\label{fig:dstst}}
\end{center}
\end{figure}

Orbitally excited (P-wave) charmed mesons have been observed in both
experiments. D0 has studied the semi leptonic decay of B mesons into these
particles and muons, measuring a branching fraction $Br(B\to \mu\nu D^{**} X)*
Br(D^{**}\to D^{*-}\pi^+ X) = (0.280\pm0.021\pm 0.088)\%$. Due to the finite
charm mass, the $D^{**}$ is actually composed of two quasi-degenerate narrow
states and two broad states; the spectrum of the two narrow states can be
seen in figure \ref{fig:dstst}.
\section*{Search for new physics}
Search of new physics in low-Pt events is either in the search for rare decays
of known states, or in the search for new states. An example of the former
is the search for the favor-changing neutral current decay $D^0\to\mu^+\mu^-$.
The standard model predicts for this decay a branching fraction of the order
of $10^{-13}$. New physics can substantially enhance this fraction, allowing
the exploration of couplings to up-type quarks not necessarily constrained by
B decays. CDF has looked for this decay into 69 $pb^{-1}$ of collected data,
observing 0 events with a background expectation of $1.6\pm 0.7$ and setting
a limit $BR(D^0\to\mu\mu)<2.5 (3.3)\times 10^{-6}$ at 90\% (95\%) C.L.,
improving by a factor 2 the present best results.\par
\begin{figure}
\begin{minipage}[h]{0.47\textwidth}
\begin{center}
\psfig{figure=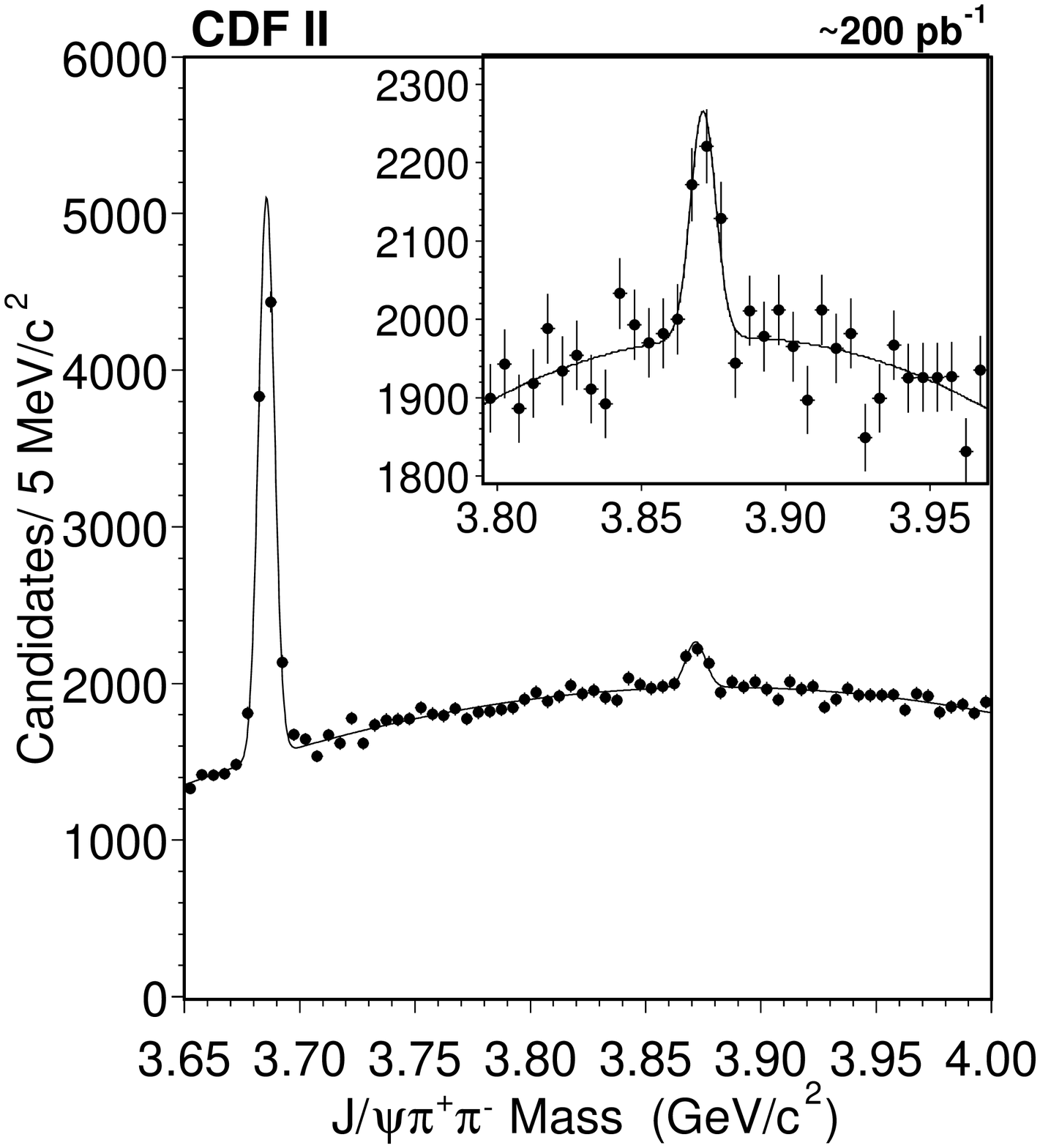,height=8cm}
\end{center}
\end{minipage}\hfill
\begin{minipage}[h]{0.47\textwidth}
\begin{center}
\psfig{figure=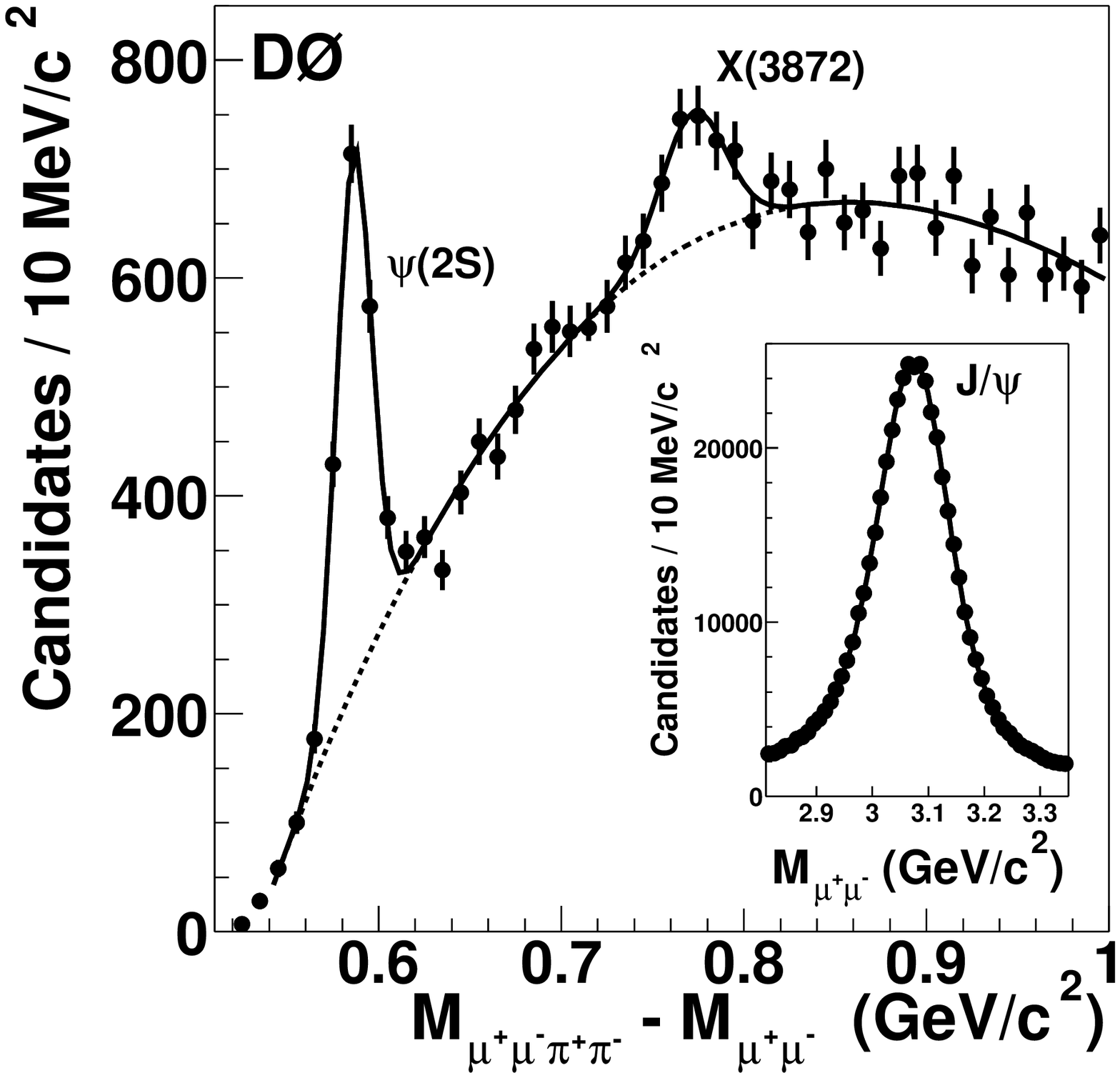,height=8cm}
\end{center}
\end{minipage}
\caption{$J/\Psi \pi^+\pi^-$ mass distribution for CDF (left) and D0 (right)
\label{fig:xmass}}
\end{figure}

The Belle collaboration has recently observed a new unexpected narrow state
X(3872) in $J/\Psi \pi\pi$ decays, with mass $M(X) = 3872 \pm 0.6 \pm 0.5$
MeV\cite{bellex}. 
This signal has been confirmed by both Tevatron detectors (figure
\ref{fig:xmass}). CDF has observed a 11$\sigma$ signal, with mass
$M(X) = 3871.3\pm 0.7\pm 0.4$ MeV, while D0 has a 4.4$\sigma$ signal, with
mass difference $M(X)-M(\Psi(2S)) = 766.4 \pm 3.5 \pm 3.9$ MeV.\par
The interpretation of this state is still under debate, with the most
likely candidates charmonium states or a $D(D^*)$ molecule, since the mass 
of this resonance is just above the sum of the masses of $D^0$ and $D^{*0}$.

\section*{Conclusions}
Despite not being dedicated to low-Pt physics, the experiments at Fermilab
play a major role in the field of charm physics, thanks to the huge production
cross section and the use of dedicated triggers. In particular, the SVT used
in CDF proved to be a huge success and opened up a full area of study of
hadronic decays of charm and beauty hadrons.

\end{document}